\documentstyle[12pt,aasms4]{article}

\received{}
\accepted{}
\journalid{}{}
\articleid{}{}
\lefthead{Hatano et al.}
\righthead{Spectra of Supernovae}

\begin{document}
\title{Ion Signatures in Supernova Spectra}

\author{Kazuhito Hatano, David Branch, Adam Fisher, Jennifer
Deaton and E. Baron}
 
\affil{Department of Physics and Astronomy, University of
Oklahoma, Norman, OK 73019; hatano, branch, fisher, deaton, baron@mail.nhn.ou.edu}

\begin{abstract} 

A systematic survey of ions that could be responsible for features in
the optical spectra of supernovae is carried out.  Six different
compositions that could be encountered in supernovae are considered.
For each composition, the LTE optical depth of one of the strongest
optical lines of each ion is plotted against temperature.  For each
ion that can realistically be considered as a candidate to produce
identifiable features in supernova spectra, a sample synthetic
spectrum is displayed.  The optical depth plots and the synthetic
spectra can provide guidance to studies of line identifications in the
optical spectra of all types of supernovae during their photospheric
phases.

\end{abstract}


\section{INTRODUCTION}

Owing to the strong Doppler broadening and line blending in supernova
spectra, making line identifications can be difficult.  Although many
of the major features in the optical spectra of supernovae have been
identified, others still lack a secure identification.  For example,
some features that have occupied our attention recently are in the
early spectra of the peculiar Type~Ia SN~1991T (Fisher et~al. 1998)
and the early spectra of the peculiar Type~Ic SN~1997ef (Deaton et~al.
1998a).  Our difficulty in identifying some of the features in the
spectra of these and other supernovae has motivated us to carry out a
systematic survey of the ions that could produce features in the
optical spectra of supernovae.  Shortly after the appearance of the
Type~II SN~1987A in the LMC, such a survey was published by Branch
(1987), but only one composition was considered: hydrogen--rich, with
a metallicity lower than solar by a factor of four.  Here we consider
six different compositions that could be encountered in supernovae.

In section 2, for each composition, LTE optical depths are calculated
in the Sobolev approximation and plotted against temperature for one
of the strongest optical lines of each ion.  The LTE approximation has
been shown to be useful for first--order interpretations of supernova
spectra (e.g., Branch et~al 1985; Jeffery \& Branch 1990; Harkness
1991a,b; Jeffery et~al 1991, 1992; Filippenko et~al. 1992a; Kirshner
et~al. 1993).  The optical depth plots of section~2 show which ions
can be considered as candidates for line identifications, and at which
temperatures.  Then a synthetic optical spectrum for each candidate
ion is displayed in section 3.  These plots show the spectral
signatures of each of the candidate ions.  By displaying sample
synthetic spectra for all of the interesting ions in this paper, we
will not have to repeat any of them in future papers based on SYNOW
calculations (e.g. Fisher et al. 1998, Deaton et al. 1998a and 1998b).

\section{LINE OPTICAL DEPTHS}

\subsection{Calculations}

In the Sobolev approximation (Sobolev 1960, Castor 1970, Jeffery \&
Branch 1990), the optical depth of a line in a supernova that is
expanding homologously with $v = r/t$ is

$$ \tau = \left(\pi e^2 \over mc\right)\ f\ \lambda\ t\ n_l\ (1 - \left(g_l
n_u \over g_u n_l\right)) \eqno (1) $$

\noindent where $n_l$ and $n_u$ are the number densities in the
lower and upper levels of the transition, $f$ is the oscillator
strength, $t$ is the time since explosion, and the other symbols have
their usual meanings.  The Sobolev line optical depth is inversely
proportional to the velocity gradient, i.e, proportional to $t$,
because as a photon propagates it redshifts with respect to the
matter; the larger the velocity gradient, the sooner the photon
redshifts out of resonance with the transition, and the smaller the
line optical depth.  A convenient numerical form of equation (1) is

$$ \tau = 0.026\ f\ \lambda_\mu\ t_d\ n_l\ (1 - \left(g_l n_u \over
g_u n_l\right)) \eqno(2)
$$

\noindent where $\lambda_\mu$ is in microns, $t_d$ is in days, and $n_l$ is in
cm$^{-3}$.

In this paper we evaluate all line optical depths at the layer in the
supernova where the electron scattering optical depth, $\tau_{es}$,
reaches unity.  When electron scattering is the dominant opacity
source, this layer can be thought of, roughly, as the bottom of the
line--forming layer (although thermalization of the continuum will
take place at a deeper layer or, in SNe Ia, perhaps not at all), 
and only transitions that achieve
optical depths on the order of unity or greater at $\tau_{es}=1$ will
be able to form conspicuous features in the spectrum.  We also will
encounter situations where the electron scattering opacity will be
smaller than the combined opacity of numerous lines. Then,
effectively, the bottom of the line forming layer will be at a
shallower layer in the supernova, and only lines that have optical
depths well in excess of unity at $\tau_{es}=1$ will be able to
produce identifiable spectral features.

If the electron density is taken to decrease outward as a power law of
index $n$, the electron density at $\tau_{es}=1$ is given by

$$ n_e = (n-1)/(\sigma_e R) \eqno (3) $$

\noindent where $\sigma_e$ is the Thomson cross section and $R$ is the
radius at which $\tau_{es}=1$.  We use $n=7$ and a characteristic
value of $R= 1.73
\times 10^{15}$ cm, which corresponds to $v_{phot}=10,000$ km s$^{-1}$ and
$t=20$ days.  This gives $n_e = 5.2 \times 10^9$ cm$^{-3}$, which is
used for all of the optical depth calculations of this paper.  A value
of $t=20$ days also is used in equation (2), except in section 2.7
where we discuss the time--dependent nickel--cobalt--iron composition
that results from the radioactive decay of $^{56}$Ni and $^{56}$Co.
The use of constant characteristic values of $n_e$ and $t$ is
sufficient for our purposes because the limiting approximation of this
work is that of LTE.

Table~1 lists the relative abundances of the elements for five of the
six compositions that we consider.  The entries are logarithms of the
abundance by number.  Only the relative abundances are relevant; the
absolute numbers have no significance.  The way in which we arrived at
each of these compositions, and where in supernovae they might be
encountered, will be discussed below as the line optical depths are
presented for each composition.  (The sixth composition that we
consider is the time--dependent mixture of nickel, cobalt, and iron
that results from an initially pure $^{56}$Ni composition.)  Given the
electron density and a composition, the Saha ionization and Boltzmann
excitation equations are used to calculate the atomic level
populations as well as the total gas density that is needed to provide
the specified electron density.  Figure 1, which shows the total
density plotted against temperature for each of the six compositions,
will be helpful in understanding the optical depth plots.  As
temperature falls, and significant donors of free electrons recombine,
the total density rises in order to achieve the specified electron
density.  For example, for the solar composition, the density rise at
$T
\la 6000$ K is caused by hydrogen recombination, and for the helium--rich
composition the steep density rise at $T \la 10,000$ K is caused by
helium recombination.

For each ion, the optical depth of a reference line is calculated as a
function of temperature.  One of the ion's strongest lines in the
range $4000 < \lambda < 10,000$ \AA\ is chosen to be the reference
line.  Note that for O~III and O~II we are using forbidden lines as
the reference lines, because they are stronger than any of the
permitted lines in the temperature range considered here.  The optical
depth plots show the reference lines that achieve $\tau > 0.001$
within the range $20,000 > T > 5,000$ K.  These reference lines are
listed in Table~2, along with their log($gf$) values and excitation
potentials.  The entry following the wavelength tells in which optical
depth plots the ion makes an appearance; for example, 234 means that
the ion appears on the optical depth plots for the second, third, and
fourth compositions that we consider.

In this paper we make no allowance for nonthermal excitation and
ionization. This is known to be significant for He~I in the
post--maximum spectra of Type~Ib supernovae (Harkness et al. 1987;
Lucy 1991) and may also be significant during certain phases of Type
II (Jeffery \& Branch 1990) and Type Ic supernovae (Clocchiatti et
al. 1997 and references therein; Deaton et al. 1998b).  Such effects
are discussed briefly below, when they may be relevant.

\subsection{Hydrogen--rich}

The early spectra of Type II supernovae are formed in a hydrogen--rich
composition, so we begin by considering the solar composition.

The optical depths are plotted in Figure 2.  At high temperature
no lines have $\tau > 1$, and only H$\alpha$ has $\tau > 0.1$.
As the temperature falls, $H\alpha$ becomes strong, followed by Ca~II,
Fe~II, Ti~II, and Sc~II.  Hydrogen is the main provider of free
electrons throughout our temperature range, but at $T
\simeq 6000$ K it begins to become mostly neutral so the
density must rise (Figure~1) to provide the specified electron
density.  This is the cause of the increase of the optical depths of
singly ionized species such as Si~II and Mg~II at $T \la 6000$ K.  At
these low temperatures, lines of many neutral species achieve $\tau >
1$, as do the resonance lines of Sr~II and Ba~II.

Because hydrogen is the main source of free electrons, the effect of
changing the metallicity is simple: the optical depths of all the
heavy--element lines are proportional to the metallicity while the
hydrogen and helium optical depths are unaffected.  Lowering the
hydrogen abundance in favor of helium would leave the hydrogen lines
almost unaffected, because the total density would rise to give almost
the same hydrogen density at the photosphere.  The density rise would
cause a strengthening of the metal lines, and helium lines would
strengthen even more because of the increased helium abundance.

Most of the line optical depths are proportional to the square of the
electron density.  One factor of $n_e$ comes just from the associated
increase in the total density, while the other comes from the Saha
equation.  This is because the lines of interest generally come not
from the most abundant ion, but from the one that is once less
ionized.  For hydrogen, for example, increasing the specified electron
density would not only increase the total hydrogen density but also
the fraction of hydrogen that is neutral, except for $T \la 6000$ K
where most of the hydrogen already is neutral.

The LTE optical depths of Figure~2 are, at least qualitatively,
nicely consistent with the observed evolution of the spectra of Type
II supernovae.  Before and around the time of maximum light, SNe~II
are hot and their spectra are continuous except perhaps for weak
hydrogen lines.  As SNe~II cool the hydrogen lines strengthen,
followed by lines of Ca~II, Fe~II, Ti~II and Sc~II, as predicted.  The
most obvious discrepancy between prediction and observation is that
the Na~I D lines are calculated to be only about as strong as the
reference lines of a number of other neutral species, whereas in the
spectra of SNe~II the feature produced by the D lines is considerably
stronger than the lines of other neutrals.  The answer, presumably,
lies in non--LTE effects.  A detailed study of line identifications in
SN 1987A was carried out by Jeffery \& Branch (1990).  In that
particular case, too, the order of appearance of the lines was very
much as discussed above.  Lines that could be identified with
confidence were from H, He~I, Na~I, Ca~II, Sc~II, Fe~II, and Ba~II.
The very early appearance of a weak He~I $\lambda$5876 line possibly 
was due to an enhanced abundance of helium in the outer hydrogen--rich
layers of SN~1987A, and the later (probable) appearance of the He~I
line is attributable to nonthermal excitation and ionization caused by
the products of the radioactive decay of $^{56}$Co.  The appearance of
conspicous Ba~II lines probably was due to an overabundance of barium
(Mazzali \& Chugai 1995 and references therein).

The satisfactory agreement between calculated LTE optical depths and
observation of SNe~II suggests that the LTE optical depths presented
below, for other compositions, also can be helpful for line identification
studies.

\subsection{Helium--Rich}

The surface layers of SNe~Ib are hydrogen--deficient and helium--rich,
as may be the surface layers of SNe~Ic.  Such a composition also will
be present, of course, beneath the outer hydrogen--rich layers of
SNe~II.  We consider a helium--rich composition that is obtained by
``burning'' all of the hydrogen to helium. This increases the helium
number abundance, relative to the heavier elements, by a factor of
3.55.  (We make no allowance for conversion of carbon and oxygen into
nitrogen by means of the CNO cycle.)

As can be seen in Figure~1, the density at the photosphere is higher
in the helium--rich case than in the hydrogen--rich case, by a factor
of about 3.1 when helium is singly ionized ($T \ga 10,000$ K), but by
a factor about 600 at $T \simeq 7500$ K, where helium is neutral.
Figure~3 shows that at low temperature, oxygen, and then carbon,
become the main sources of free electrons.

The line optical depths are shown in Figure~4.  Apart from the
absence of the hydrogen line and the strengthening of He~I and He~II
due to the abundance change, the differences between the optical
depths in the helium--rich and hydrogen--rich cases just reflect the
difference in the density at $\tau_{es}=1$, as a function of
temperature. At high temperature no lines achieve $\tau > 1$, and the
He~I line just makes it at $T \simeq 8000$ K.  Considering the
limitations of the LTE approximation, this means that the absence of
conspicuous He~I lines, at any temperature, is not evidence against a
helium--rich composition. For $T \la 10,000$ K, helium begins to
recombine, the density rises sharply, the lines of heavy elements
become much stronger at $\tau_{es}=1$ than they were in the
hydrogen--rich case, and line opacity dominates over electron
scattering.

As is well known, the strong He~I lines that are observed in the
post--maximum light photospheric phases of SNe~Ib require nonthermal
excitation and ionization.  Because the nonthermal effects increase
the ionization of helium, whenever they occur they will cause the
sharp density rise and the corresponding strengthening of the heavy
element lines to shift to lower temperature than seen in Figure~4.

\subsection{Carbon/Oxygen--Rich}

The outermost layers of SNe~Ia and SNe~Ic may be carbon/oxygen--rich.
For example, in the model W7 (Nomoto, Thielemann, \& Yokoi 1984) the
composition is C/O--rich down to 14,900 km s$^{-1}$. We consider a
composition that is obtained by burning all of the helium into equal
amounts of carbon and oxygen by mass.  This raises the carbon and
oxygen number abundances, relative to the heavy elements, by factors
of 147 and 52, respectively, and makes carbon more abundant than
oxygen by a factor of 1.3.

Carbon and oxygen are the principal sources of free electrons.
Figure~1 shows that the density does not vary strongly with
temperature.  For $T \ga 10,000$ K the density is somewhat higher than
in the hydrogen and helium rich cases, but the density does not
increase sharply at low temperature because carbon remains partially
ionized down to 5000~K.

The line optical depths are shown in Figure~5.  Now, for the
first time, we see a line having $\tau > 1$ at high temperature ---
that of C~III --- and at intermediate temperature the C~II line has
$\tau > 1$.  The forbidden reference lines of [O III] and [O II] have
$\tau > 0.1$.  At low temperature, the O~I and C~I lines are only
about as strong as they were in the helium--rich case, because in both
cases the free electrons are mainly provided by carbon and oxygen.
This means that, at least in LTE, the strength of the O~I
$\lambda$7773 line cannot discriminate between a helium--rich and a
C/O--rich composition.  At low temperature the lines of elements
heavier than oxygen are stronger than they were in the hydrogen--rich
case, but not nearly as strong as they were in the helium--rich case,
because their abundances relative to the main providers of free
electrons, carbon and oxygen, are now reduced.

\subsection{Carbon--burned}

Layers in which carbon has burned are expected to be encountered in
the surface or subsurface layers of SNe~Ia, and they are exposed above
the photosphere before the time of maximum light. In model W7, the
carbon--burned composition extends from 14,900 to about 13,000 km
s$^{-1}$.  For a carbon--burned zone we consider composition number 5
from Table~3 of Khokhlov, M\"uller, \& H\"oflich (1993).  Oxygen is
the most abundant element, followed by silicon, sulfur, and magnesium.
Oxygen is the main source of free electrons, except for $T \la 6000$ K
where it recombines and silicon takes over.  Figure~1 shows that the
density runs somewhat higher than that of the C/O--rich composition.

The optical depths are plotted in Figure~6.  At high and
intermediate temperatures, lines of silicon and sulfur ions are
strong, and it is interesting that lines of P~III and P~II also have
$\tau > 1$.  At $T \la 8000$ K lines of Si~II, Ca~II, and Mg~II become
very strong, and at $T
\la 7000$ K lines of additional singly ionized and some neutral species also
become strong.  The O~I line does not get much stronger than it was in
the helium--rich and C/O--rich cases, so again the O~I $\lambda$7773
line is not a good composition indicator.

\subsection{Oxygen--burned} 

In model W7 a layer in which oxygen has burned extends from about 9000
to 13,000 km s$^{-1}$.  We consider composition number 4 from Table~3
of Khokhlov et~al. (1993).  The most abundant elements are silicon,
sulfur, iron, argon, and calcium.  Silicon is the main source of free
electrons, and the density runs somewhat higher than in the
carbon--burned case (Figure~1).

The optical depths are plotted in Figure~7. The main qualitative
differences between these plots and those of the carbon--burned case
are the disappearance of the oxygen and phosphorus lines, the
weakening of Mg~II, and the strengthening of Ca II, Fe~III, and Fe~II.

\subsection{Nickel--decay}

The sixth composition that we consider is the time dependent mixture
of nickel, cobalt, and iron that results from the decay of $^{56}$Ni
through $^{56}$Co to stable $^{56}$Fe.  This composition has been
suggested to dominate the outermost layers of the peculiar Type Ia SN
1991T (Filippenko et~al. 1992b, Ruiz--Lapuente et~al. 1992), or at
least to be an significant component of those layers (Jeffery et
al. 1992, Mazzali et al. 1995, Fisher et al. 1998), and large 
amounts of $^{56}$Ni are
present beneath the oxygen--burned zone in normal SNe~Ia.  Figure~8 shows
the time dependent fractional abundances of nickel, cobalt, and iron,
assuming an initially pure $^{56}$Ni composition and using half lives
of 6 and 77 days for $^{56}$Ni and $^{56}$Co, respectively.

The density plotted in Figure~1 is for $t = 20$ days. It is only a
very weak function of the time, owing to the similar ionization
potentials of nickel, cobalt, and iron.  The density runs somewhat
higher than in the oxygen--burned case, and it does not vary strongly
with temperature.

Figure~9 shows the line optical depths for three different times:
$t=5$ days, when the fractional concentrations of nickel, cobalt, and
iron are 0.55, 0.45, and $7 \times 10^{-3}$, respectively; at $t=20$
days (0.09, 0.81, 0.1); and $t=80$ days ($7 \times 10^{-5}$, 0.53,
0.47).  At 5 and 20 days, lines of the doubly or (depending on
temperature) singly ionized species of all three elements are
candidates for producing spectral features.  At 80 days nickel lines
are no longer a consideration.

\section{SYNTHETIC SPECTRA}

In Figure~10 a sample synthetic spectrum is displayed for every ion
that we consider to be a realistic candidate for producing an
individually recognizable feature in the optical spectrum of a
supernova. Not every ion that is listed in Table~2 appears in Figure
10. An ion such as V~I, for example, that appears on an optical depth
plot only in situations where numerous other ions are much stronger,
is not a realistic candidate for producing identifiable spectral
features. 

Although the reference lines were selected from lines that have
$\lambda < 10,000 $ \AA, the synthetic spectra of Figure~10 extend to
12,000 \AA.  The purpose of this is to show the feature produced by
$\lambda$10830 of He~I, as well as features produced by other species
such as Si~I, which might produce the near infrared absorption feature
that was observed in the spectrum of the Type Ic SN~1994I by
Filippenko (1995).  A discussion of the identification of this feature
is presented by Deaton et~al. (1998b).

The synthetic spectra are calculated with the parameterized supernova
spectrum synthesis code SYNOW, which is an improved version of the
code that has been used by in the past by, for example, Branch
et~al. (1985) and Filippenko et~al. (1992a).  The current version of
SYNOW is described briefly by Fisher et~al. (1997) and in detail by 
Fisher (1999).  For the purposes
of this spectrum atlas, we use the following parameters: the
temperature of the blackbody continuum is 10,000 K; the line optical
depths decrease outwards as a power law of index 7; the velocity at
the photosphere is 5000 km s$^{-1}$; and the line forming region is
truncated on the outside at a velocity of 30,000 km s$^{-1}$.  The
optical depth of the reference line at the photosphere ordinarily is
taken to be 10, but for some ions it is raised or lowered in order to
better illustrate the spectral signature of the ion.  Different
excitation temperatures are used for different ionization stages
because, for example, lines of most neutral species have no chance of
being seen unless the temperature is low.  The excitation temperature
is taken to be 5000 K for neutral species; 10,000 K for singly
ionized; 15,000 K for doubly ionized; and 20,000 K for triply ionized.

Our intention here has been to show a synthetic spectrum for every ion
that can realistically be considered as a candidate for producing an
identifiable feature in the optical spectra of supernovae.
Consequently the optical depth plots and synthetic spectra displayed
in this paper can be used as a guide to line identifications in the
optical spectra of all types of supernovae in their photospheric
phases.

Figures and electronic versions of figures 10 may be obtained at:  http://www.nhn.ou.edu/$\sim$baron/papers.html

\acknowledgments 

This work has been supported by NSF grants AST 9417102, 9417242, and 9731450.

\clearpage

\begin{figure}
\figcaption{The log of the total density (gm cm$^{-3}$) at $\tau_{es}=1$
is plotted against temperature for six different compositions.
\label{fig1}} \end{figure}

\begin{figure}
\figcaption{ The log of the Sobolev LTE optical depth of ion reference
lines, evaluated at $\tau_{es}=1$, is plotted against temperature for
the hydrogen--rich composition.
\label{fig2}} \end{figure}

\begin{figure}
\figcaption{The fractional contribution of free electrons is plotted
against temperature for the helium--rich composition.
\label{fig3}} \end{figure}

\begin{figure}
\figcaption{The log of the Sobolev LTE optical depth of ion reference
lines, evaluated at $\tau_{es}=1$, is plotted against temperature for
the helium--rich composition.
\label{fig4}} \end{figure}

\begin{figure}
\figcaption{The log of the Sobolev LTE optical depth of ion reference
lines, evaluated at $\tau_{es}=1$, is plotted against temperature for
the C/O--rich composition.
\label{fig5}} \end{figure}

\begin{figure}
\figcaption{The log of the Sobolev LTE optical depth of ion reference
lines, evaluated at $\tau_{es}=1$, is plotted against temperature for
the carbon--burned composition.
\label{fig6}} \end{figure}

\begin{figure}
\figcaption{The log of the Sobolev LTE optical depth of ion reference
lines, evaluated at $\tau_{es}=1$, is plotted against temperature for
the oxygen--burned composition.
\label{fig7}} \end{figure}

\begin{figure}
\figcaption{The fractional composition is plotted against time for the
$^{56}$Ni--decay composition.
\label{fig8}} \end{figure}

\begin{figure}
\figcaption{The log of the Sobolev LTE optical depth of ion reference
lines, evaluated at $\tau_{es}=1$, is plotted against temperature for
the $^{56}$Ni--decay composition at times of 5, 20, and 80 days.
\label{fig9}} \end{figure}

\begin{figure}
\figcaption{A synthetic spectrum is shown for each ion that can be
regarded as a candidate for producing identifiable features in the
photospheric--phase spectra of supernovae.
\label{fig10}} \end{figure}


\clearpage

\begin{references}

\reference{} Branch, D. 1987, ApJ, 320, L121

\reference{} Branch, D., Doggett, J. B., Nomoto, K., \& Thielemann,
F.--K. 1985, ApJ, 294, 619

\reference{} Castor, J. I. 1970, MNRAS, 149, 111

\reference{} Clocchiatti, A., et al. 1997, ApJ, 483, 675 

\reference{} Deaton, J., Branch, D., Fisher, A., \& Baron, E. 1998a,
in preparation

\reference{} Deaton, J. et al. 1998b, in preparation 

\reference{} Filippenko, A. V., et al. 1992a, AJ, 104, 1543 

\reference{} Filippenko, A. V., et al. 1992b, ApJ, 384, L15

\reference{} Filippenko, A. V., et al. 1995, ApJ, 450, L11

\reference{} Fisher, A. 1999, Ph.D. thesis, University of Oklahoma

\reference{} Fisher, A., Branch, D., Nugent, P., \& Baron, E. 1997,
ApJ, 481, L89

\reference{} Fisher, A., Branch, D., Hatano, K., \& Baron, E. 1998,
MNRAS submitted

\reference{} Harkness, R. P. 1991a, in Supernovae, ed. S. E. Woosley
(New York, Springer), 454

\reference{} Harkness, R. P. 1991b, in SN1987A and Other Supernovae,
ed. I. J. Danziger \& K. Kj\"ar (Garching, ESO), 447

\reference{} Harkness, R. P., et al. 1987, ApJ, 317, 355

\reference{} Jeffery, D. J., Leibundgut, B., Kirshner, R. P., Benetti,
S., Branch, D., \& Sonneborn, G. 1992, ApJ, 397, 304

\reference{} Jeffery, D. J., \& Branch, D. 1990, in Jerusalem Winter
School for Theoretical Physics, Vol. 6, Supernovae, ed. J. C. Wheeler,
T. Piran, \& S. Weinberg (Singapore, World Scientific), 90

\reference{} Jeffery, D. J., Branch, D., Filippenko, A. V., \& Nomoto,
K. 1991, ApJ, 377, L89

\reference{} Khokhlov, A. J., M\"uller, E., \& H\"oflich, P. 1993,
A\&A, 270, 223

\reference{} Kirshner, R. P., et al. 1993, ApJ, 415, 589

\reference{} Lucy, L. B. 1991, ApJ, 383, 308

\reference{} Mazzali, P. A., \& Chugai, N. 1995, A\&A, 303, 118

\reference{} Mazzali, P. A., Danziger, I. J., \& Turatto, M. 1995,
A\&A, 297, 509 

\reference{} Nomoto, K., Thielemann, F.--K., \& Yokoi, K. 1984, ApJ,
286, 644

\reference{} Ruiz--Lapuente, P., et al. 1992, ApJ, 387, L33

\reference{} Sobolev, V. V. 1960, Moving Envelopes of Stars
(Cambridge, Harvard University)

\end{references}
\end{document}